\documentclass[a4paper,12pt,reqno]{amsart}
\usepackage{bookmark}

\usepackage{amsmath,amssymb}
\usepackage{geometry}
\usepackage{enumerate}
\usepackage{ifthen}
\usepackage{multirow}
\usepackage{calc}
\usepackage{mathtools}
\usepackage{multicol}
\DeclarePairedDelimiter\ceil{\lceil}{\rceil}
\DeclarePairedDelimiter\floor{\lfloor}{\rfloor}
\nonstopmode\numberwithin{equation}{section}
\usepackage{amssymb,latexsym}
\usepackage{graphicx}

\newtheorem{definition}{Definition}
\newtheorem{example}{Example}
\allowdisplaybreaks
\usepackage{tikz}
\usetikzlibrary{matrix,shapes,arrows,positioning,chains}
\usetikzlibrary{decorations.text}
\usetikzlibrary{decorations.pathmorphing}

\begin{document}
\title{Non-Singular Transformation Based Encryption Scheme}
\maketitle
 \begin{center}
\center{  $\text{Sunil Kumar}^1, \text{Sandeep Kumar}^1, \text{Gaurav Mittal}^{1,2}, \text{Shiv Narain}^3 $ }
{\footnotesize

 \center{$^1$DRDO, India \\$^{2}$Department of Mathematics, Indian Institute of Technology Roorkee, Roorkee, India\\ $^{3}$Department of Mathematics Arya P.G. College Panipat, India\\ emails: sunilsangwan6174@gmail.com, sandeepkumar@hqr.drdo.in\\  gmittal@ma.iitr.ac.in, math.shiv@gmail.com
 }
 }
 \end{center}
 \medskip
\begin{abstract}
	In this paper, we propose a novel variant of the Hill cipher based on vector spaces. In the classical Hill cipher, a non-singular matrix is used for encryption but it is well known that this cipher is vulnerable to the known-plaintext attack. In our proposed cryptosystem, we eradicate this problem by encrypting each plaintext block with a new invertible key matrix. To generate the invertible matrices which serve as the dynamic keys, we make use of the vector spaces along with randomly generated basis and non-singular linear transformation. In addition to this, we also study  the computational complexity of the proposed cryptosystem and compare this with the  computational complexities of other schemes based on Hill cipher.
\end{abstract}
\begin{center}
	\keywords {\hspace{-7mm}\textbf{Keywords:}  Hill cipher, Vector space, Basis, Symmetric Cryptography, Non-Singular Transformation}\\ 
	\subjclass{\hspace{-06mm} \textbf{Mathematics Subject Classification(MSC) (2010)}: 11T71, 94A60}
\end{center}
\section{Introduction}
There has been the requirement of secure communication since thousands of years which led the way for invention of cryptography. 
Cryptography enables two persons, let's say, Alice and Bob to  communicate securely assuming the fact that channel between them is monitored by an adversary. Secure communication can be done by  using the symmetric key cryptography   or asymmetric key cryptography  or combination of both known as  hybrid cryptography \cite{katz1996handbook}. In this paper, we restrict ourselves to symmetric key cryptography (in particular, Hill cipher) which involves prior sharing of a secret key between Alice and Bob.  Hill cipher \cite{mahmoud2014hill,almutairi2019new}is one of the oldest known polyalphabetic cipher invented by Lester S. Hill \cite{hill1929cryptography},   but it is well known that various attacks can be implemented on this cipher  (e.g. known plaintext attack)\cite{stinson2005cryptography}. Despite  all these vulnerabilities, it is still gaining the attention of many researchers because of its simplicity, high speed, high throughput \cite{toorani2009secure}, ability to disguise the letter frequencies and importance in educational systems. 

Now, let us mention some of the literature available in this direction.
Yeh et al. \cite{yeh1991new} proposed an improvement of the Hill cipher by presenting a polygraph substitution algorithm. Although their algorithm is safe against the known-plaintext attack, but as discussed in \cite{reddy2012modified}, it is inefficient for bulk data and is time-consuming. Sadeenia \cite{saeednia2000make}  gave a way for enhancing the security of the Hill cipher by randomly  permuting the rows and columns of a master key matrix and use them as a dynamic key matrix. In this scheme, both the encrypted permutation vector and encrypted plaintext are transmitted at the receiver end. Clearly,  each plaintext is encrypted with the help of a new matrix but the known-plaintext attack  can still be applied on the permutation vector \cite{lin2004comments}, i.e. same vulnerability as in the Hill cipher. A modification of Hill cipher  similar to \cite{saeednia2000make} is proposed in \cite{chefranov2008secure}. This modification involves the use of a pseudo-random permutation generator by both the sides and sharing of necessary  permutations  with the   receiver. Ismail et al. \cite{ismail2006repair} gave \textit{HillMRIV}, which is a variant  of the Hill cipher. This variant uses a new key matrix for encrypting each plaintext block rather than using a single key matrix for the entire plaintext. This increases the security of the Hill cipher by thwarting the known-plaintext attack, but the encryption scheme has a severe issue regarding the invertibility of the key matrix \cite{li2008cryptanalysis} which is nowhere discussed in the paper. Mahmoud \cite{mahmoud2009hill} gave a modification of the Hill cipher based on the generation of dynamic encryption key matrix by exponentiation with the help of eigenvalues. Further, circulant matrices  have been employed in \cite{reddy2012modified} to give another variant of Hill cipher  and claimed to be secure against the chosen-plaintext as well as known-plaintext attack. But, in \cite{ElHabshy2019} it has been shown that it is vulnerable to both the attacks. Acharya et al. \cite{acharya2009image}  proposed an algorithm termed as AdvHill which make use of the involutory key matrix for both encryption and decryption. Since an involutory matrix is self invertible, so the computation involved in finding inverse of the key matrix is not required in this algorithm. Agrawal et al. \cite{agrawal2014elliptic} discussed a new encryption method  in which the Hill cipher is generated with elliptic curves. This method increases the security but is inefficient because of the geometric structure of the elliptic curves.
\\ 

\textbf{Organisation and Contribution:} In this paper, we introduce a novel variant of the Hill cipher  that conquers all the security snags. More specifically, we propose a vector space-based approach in which dynamic keys are produced with the help of a  non-singular transformation and basis. We also show that the known-plaintext attack is no more a vulnerability in our proposed scheme. We also study the bit operation complexities of our scheme and make a comparison of the bit operations required for other variants of Hill cipher. The rest of the paper is organized as follows: All the necessary definitions are presented in Section $2$. Section $3$ is devoted to the our novel scheme. Much required security analysis and computational costs of the scheme is discussed in Section $4$ and $5$,  respectively. A toy example for the feel of the scheme is given in the last section.

\section{Preliminaries}

\begin{definition}Linear Transformation: 
	Let $V_1$ and $V_2$ be two vector spaces over the field $\mathbb{F}$. A map $T:V_1\to V_2$ is termed as a linear
	transformation provided $$T(c_1\alpha_1+c_2\alpha_2) = c_1T(\alpha_1)+c_2T(\alpha_2)$$
	for all $\alpha_1,\alpha_2 \in V$ and  $c_1,c_2\in \mathbb{F}$.
\end{definition}
\begin{definition}Non-singular Linear Transformation:  
	A linear transformation $T:V\to V$ is   non-singular, if for any $v\in V$, $T(v) = 0$ implies $v = 0$. 
\end{definition}
Finally, let us end this section by recalling the well known classical Hill Cipher in the following subsection.
\subsection{The Hill Cipher} In this cipher, before encryption, the plaintext is written into blocks of equal length say $m$. Then  each plaintext block is encrypted using an invertible matrix of size $m$ to obtain the corresponding  ciphertext block of the same length.  Decryption can be done simply by using the inverse of the matrix used for encryption. Mathematically, 
encryption and decryption process in the Hill cipher are as follows:
Define  $\mathcal{P}=\mathcal{C}=(\mathbb{Z}_{26})^m$ which means digits of the message and ciphertext are ranging from $0$ to $25$. 
Let
$$\mathcal{K}=\{K\ |\ K\   \text{is a  inverible square matrix of order}\  m \text{ over}\ \mathbb{Z}_{26} \}.$$
For a key $K \in \mathcal{K}$,  define the encryption and decryption functions $$E:\mathcal{K}\times \mathcal{P}\to \mathcal{C}: E(K, \mathrm{x})= \mathrm{x}K,$$ $$D:\mathcal{K}\times \mathcal{C}\to \mathcal{P}: D(K, \mathrm{y})= \mathrm{y}K^{-1},$$ respectively, 
where all the calculations are performed in $\mathbb{Z}_{26}$.

To make the decryption possible, it is important that the key matrix $K$ must be invertible. But every square matrix over $\mathbb{Z}_{26}$ is not invertible. Further, if we consider the ring $\mathbb{Z}_{p}$, for some prime $p$ instead of $\mathbb{Z}_{26}$, then this increases the probability of randomly choosing a matrix  whose determinant is non-unit. This selection of ring also increases the size of key space \cite{overbey2005keyspace}. 
If the adversary is available with ciphertexts only, then breaking the Hill cipher is an uphill task. However, the known plaintext attack still makes it most vulnerable \cite{stinson2005cryptography}. To see this,  suppose the attacker is in the possession of at least $m$ different pairs $x_{s}=(x_{1s},x_{2s},x_{3s}\cdots,x_{ms})$ and $y_{s}=(y_{1s},y_{2s},y_{3s}\cdots,y_{ms})$, where $y_{s}$ are encryptions of $x_{s}$ for $1\leq s\leq m.$ Using two square matrices $X=(x_{rs})$ and $Y=(y_{rs}),$ a matrix equation $Y=XK$ can be formed, where the unknown square matrix $K$ of order $m$ is the key matrix. If $X$ has inverse, then by  premultiplying $Y=XK$ by $X^{-1}$, $K$ can be recovered and the system will be broken. If $X$ does not have inverse, then try another $m$ plaintext-ciphertext pairs.  This shows that  the known plaintext attack can be implemented on the Hill cipher.

\section{The Proposed Scheme} In this section, we introduce the novel variant of the Hill cipher based on vector spaces. 
Let $p$ be a large prime and $\mathbb{F}_p$ be the corresponding field. Further let $$V= \big\{(x_1, x_2, \cdots, x_n):x_i \in \mathbb{F}_p,\ \text{for}\  1\leq i \leq n\big\},$$  be a $n-$dimensional vector space over the field $\mathbb{F}_p$ and  $T:V \to V$ be a non-singular transformation. Since every non-singular linear transformation corresponds to a non-singular matrix over the field $\mathbb{F}_p,\ T$ can be seen as an element of $ GL(n,\mathbb{F}_p)$, which is a group of all non-singular matrices of order $n$ over the field $\mathbb{F}_p.$ We choose one such $T$ whose order is large in $ GL(n,\mathbb{F}_p)$. Rationale behind choosing this $T$ is discussed later  on. 

Basically, our scheme involves the following main points which will be discussed in the following  subsections:\begin{enumerate}
	\item Construction of a linear transformation $T$.
	\item How to write the plaintext in blocks?
	\item Whitening process before encryption.
	\item How to generate the dynamic keys?
	\item Encryption and decryption algorithms.
\end{enumerate}
To start with, first of all, we discuss how to select a non-singular linear transformation.
\subsection{How to construct $T$}
It is well known that the vector space of all square matrices of order $n$ (i.e. $M(n,\mathbb{F})$ over a field $\mathbb{F}$ is isomorphic to the vector space Hom$_\mathbb{F}(V,V)$ of all linear transformations on an $n$-dimensional vector space $V$ over $\mathbb{\mathbb{F}}$ \cite{hoffmanlinear}. Mathematically, we can write $$M(n,\mathbb{F})\cong \ \text{Hom}_\mathbb{F}(V,V).$$ This means, any invertible matrix in $M(n,\mathbb{F})$ can be seen as a non-singular linear transformation.
\begin{example}
	Let $A=\begin{bmatrix}
	4&2\\0&3
	\end{bmatrix}$ be a $2\times 2$ non-singular matrix over the  field $\mathbb{F}_5$. Then the transformation $T$ on $2-$dimensional vector space $V$ defined by$$ T(a, b)=\begin{bmatrix}
	a&b
	\end{bmatrix} \begin{bmatrix}
	4&2\\0&3
	\end{bmatrix}=(4a, 2a+3b), $$ is an invertible transformation.
	
\end{example}
\subsection{Way of writing the plaintext}

Write the plaintext message in blocks of length $n$,  where $n$ is some positive integer (add padding if required), i.e. $m=m_1m_2\cdots$, where $m_i= (m_i^1, m_i^2, \cdots, m_i^n)$ is $i^{\text{th}}$ message block of length $n$ with ${m}_i^j \in \mathbb{F}_p$ for all $1\leq j\leq n$.
\subsection{Whitening process}   
Choose a random $1\times n$ non zero vector  $I_1= (I_1^1, I_1^2, \cdots, I_1^n)$ with $I_1^i \in \mathbb{F}_p, 1\leq i \leq n$. This vector is  used for whitening of the first message block. For this, simply add $I_1$ and $m_1$ modulo $p$. Let $$m_1'= m_1\oplus_pI_1.$$ For the whitening of subsequent blocks, we make the use of the non-singular linear transformation  $T$ in the following manner: Let $m_i$ be the $i^{\text{th}},\ i \geq 2$ message block and $$I_i= T(I_{i-1})= (I_i^1, I_i^2, \cdots, I_i^n)\pmod p, \ \ \text{for}\ \ i\geq 2,$$ i.e. $I_i$ is obtained from $I_{i-1}$ by applying $T$ on it. Now the $i^{\text{th}}, i\geq 2$ whitened message block is $$m'_i= m_i\oplus_pI_i.$$ This process is included as  a part of encryption to overcome the problem of a message block with all entries $0$.{\color{red}{ Fig. $1$,}} represents this process schematically,  where $\oplus_p$ is the same as the addition modulo $p$.
\begin{figure}
	
	Fig. {1:}{ Whitening Process for a Message of Three Blocks}
	\tikzset{
		desicion/.style={
			diamond,
			draw,
			text width=3em,
			text badly centered,
			inner sep=0pt
		},
		block/.style={
			rectangle,  rounded corners=10pt, text width=1.75cm, 
			minimum height=1.5cm,
			draw,solid,line width=.2mm,
			rounded corners
		},
		cloud/.style={
			draw,
			ellipse,
			minimum height=2em
		},
		descr/.style={
			fill=white,
			inner sep=2.5pt
		},
		connector/.style={
			-latex,
			font=\scriptsize
		},
		rectangle connector/.style={
			connector,
			to path={(\tikztostart) -- ++(#1,0pt) \tikztonodes |- (\tikztotarget) },
			pos=0.3
		},
		rectangle connector/.default=-2cm,
		straight connector/.style={
			connector,
			to path=--(\tikztotarget) \tikztonodes
		}
	}
	\begin{tikzpicture}\hspace{.1cm}
	
	\matrix (m)[matrix of nodes, column  sep=.6cm,row  sep=3mm, align=center, nodes={rectangle,draw, anchor=center} ]{
		\\
		&  |[block]|{$m_{1}$}        &      |[block]|{$m_{1}\oplus_{p}I_{1}=m_{1}^{\prime}$}  & |[block]|{$I_{1}$}  &   \\
		|[block]|{$m=m_{1}m_{2}m_{3}$} &|[block]|{$m_{2}$} & |[block]|{$m_{2}\oplus_{p}I_{2}=m_{2}^{\prime}$} &|[block]|{$I_{2}$}  &    |[block]| {$I_{1},T(I_{1})=I_{2},T(I_{2})=I_{3}$ }                                         \\
		&  |[block]|{$m_{3}$} &|[block]|{$m_{3}\oplus_{p}I_{3}=m_{3}^{\prime}$}  &|[block]| {$I_{3}$ }  &                                        \\                                        
	};
	\draw [line width=0.2mm,>=latex,->] (m-3-1) -- (m-3-2);
	\draw [line width=0.2mm,>=latex,->] (m-3-1) -- (m-2-2);
	\draw [line width=0.2mm,>=latex,->] (m-3-1) -- (m-4-2);
	\path [line width=0.2mm,>=latex,->] (m-3-1) edge (m-3-2);
	\path [line width=0.2mm,>=latex,->] (m-2-2) edge (m-2-3);
	\path [line width=0.2mm,>=latex,->] (m-2-4) edge (m-2-3);
	\path [line width=0.2mm,>=latex,->] (m-3-2) edge (m-3-3);
	\path [line width=0.2mm,>=latex,->] (m-3-4) edge (m-3-3);
	\path [line width=0.2mm,>=latex,->] (m-4-2) edge (m-4-3);
	\path [line width=0.2mm,>=latex,->] (m-4-4) edge (m-4-3);
	\draw [line width=0.2mm,>=latex,->] (m-3-5) -- (m-2-4);
	\draw [line width=0.2mm,>=latex,->] (m-3-5) -- (m-3-4);
	\draw [line width=0.2mm,>=latex,->] (m-3-5) -- (m-4-4);
	
	\end{tikzpicture}
\end{figure}
\subsection{Key Generation Scheme}
Choose a random basis of $V$. For that, we need a non-singular square matrix over $\mathbb{F}_p$. Let $P$ be the probability of a randomly selected square matrix of order $n$ to be non-singular over modulo $p$. Then We have \begin{equation*}
P=\frac{(p^n-1)(p^n-p)\cdots (p^n-p^{n-1})}{p^{n^2}}=\frac{p^{n^2}(1-\frac{1}{p^n})(1-\frac{1}{p^{n-1}})\cdots (1-\frac{1}{p})}{p^{n^2}}
\end{equation*} 
\begin{equation*}
=(1-\frac{1}{p^n})(1-\frac{1}{p^{n-1}})\cdots (1-\frac{1}{p})
\end{equation*}
which is approximately $1$ for large $p$ and finite $n$. Therefore, any randomly selected  square matrix of order $n$ over $\mathbb{F}_p$ is probably non-singular (i.e. invertible). If not, choose another random matrix.   Within a few choices, we get the non-singular matrix. So, let $A_1$ be the chosen non-singular matrix, i.e. $A_1 \in GL(n,\mathbb{F}_p)$. This is the key matrix for encrypting first whitened block and the  set of all  rows (let's say $B_1)$ of $A_1$  serves the purpose of a random basis.

Since $T$ is a non-singular linear transformation, it maps basis to basis \cite{hoffmanlinear} and therefore by giving a random basis as a seed, it will generate a sequence of basis $\{B_2, B_3, \cdots,\}$. From this sequence of basis, we obtain a sequence of matrices in $GL(n,\mathbb{F}_p)$ by putting the elements of $B_i$ as the rows of $A_i$ for $i\geq2$ and these matrices serve as dynamic keys for the proposed algorithm. {\color{red}Fig. $2$,}  represents this scheme schematically.
\subsection{Encryption Algorithm}
First, encrypt $m_1'$ with the key matrix $A_1$ by simply multiplying  both modulo $p$. This yields ciphertext $$c_1=m_1'\otimes_p A_1.$$ 
For encrypting $i^{\text{th}}, i\geq 2$ message block, we use the key matrix $A_i$ defined in Subsection $3.4$.  Corresponding ciphertext is $$c_i=m_i'\otimes_p A_i.$$ Fig. $3$, represents this scheme schematically, where $\otimes_p$ represents    multiplication modulo $p$. 
\subsection{Symmetric key}
Since Hill cipher is a symmetric encryption algorithm, the following three things need to be shared with the decryptor in the beginning:
\begin{enumerate}
	\item Initial vector.
	\item Non-singular linear transformation.
	\item Basis of the vector space.
\end{enumerate}
\begin{figure}
	Fig.{ 2:} {Key Generation Process for a Message of Three Blocks}\vspace{.6cm}
	\tikzset{
		desicion/.style={
			diamond,
			draw,
			text width=3em,
			text badly centered,
			inner sep=0pt
		},
		block/.style={
			rectangle,  rounded corners=10pt, text width=1.62cm, 
			minimum height=1.5cm,
			draw,solid,line width=.2mm,
			text centered,
			rounded corners
		},
		cloud/.style={
			draw,
			ellipse,
			minimum height=2em
		},
		descr/.style={
			fill=white,
			inner sep=2.5pt
		},
		connector/.style={
			-latex,
			font=\scriptsize
		},
		rectangle connector/.style={
			connector,
			to path={(\tikztostart) -- ++(#1,0pt) \tikztonodes |- (\tikztotarget) },
			pos=0.3
		},
		rectangle connector/.default=-2cm,
		straight connector/.style={
			connector,
			to path=--(\tikztotarget) \tikztonodes
		}
	}
	\begin{tikzpicture}\hspace{2mm}
	
	\matrix (m)[matrix of nodes, column  sep=.8cm,row  sep=.3cm, align=center, nodes={rectangle,draw, anchor=center} ]{
		& &  |[block]|{$GL(n,p)$}	\\
		& |[block]|{$A_{1}$}        &     & |[block]|{$B_{1}, ~T$}     \\
		&|[block]|{$A_{2}$} & & |[block]|{$T(B_{1})=B_{2}$}                                         \\
		& |[block]|{$A_{3}$} & &|[block]|{$T(B_{2})=B_{3}$}                                          \\                                        
	};
	\draw [line width=0.2mm,>=latex,->] (m-1-3) -- (m-2-2);
	\draw [line width=0.2mm,>=latex,->] (m-1-3) -- (m-2-4);
	\draw [line width=0.2mm,>=latex,->] (m-2-2) -- (m-2-4);
	\draw [line width=0.2mm,>=latex,->] (m-2-4) -- (m-3-4);
	\draw [line width=0.2mm,>=latex,->] (m-3-4) -- (m-4-4);
	\draw [line width=0.2mm,>=latex,->] (m-4-4) -- (m-4-2);
	\draw [line width=0.2mm,>=latex,<-] (m-3-2) -- (m-3-4);
	\end{tikzpicture}
\end{figure}
\begin{figure}
	Fig.{ 3:}
	{Encryption Process for a  Message of Three Blocks}
	\tikzset{
		desicion/.style={
			diamond,
			draw,
			text width=3em,
			text badly centered,
			inner sep=0pt
		},
		block/.style={
			rectangle,  rounded corners=10pt, text width=1.62cm, 
			minimum height=1.5cm,
			draw,solid,line width=.2mm,
			text centered,
			rounded corners
		},
		cloud/.style={
			draw,
			ellipse,
			minimum height=2em
		},
		descr/.style={
			fill=white,
			inner sep=2.5pt
		},
		connector/.style={
			-latex,
			font=\scriptsize
		},
		rectangle connector/.style={
			connector,
			to path={(\tikztostart) -- ++(#1,0pt) \tikztonodes |- (\tikztotarget) },
			pos=0.3
		},
		rectangle connector/.default=-2cm,
		straight connector/.style={
			connector,
			to path=--(\tikztotarget) \tikztonodes
		}
	}
	\begin{tikzpicture}\hspace{2mm}
	
	\matrix (m)[matrix of nodes, column  sep=.7cm,row  sep=3mm, align=center, nodes={rectangle,draw, anchor=center} ]{
		\\
		&  |[block]|{$m_{1}'$}        &      |[block]|{$m_{1}'\otimes_{p}A_{1}=c_{1}$}  & |[block]|{$A_{1}$}  &   \\
		|[block]|{$m'=m_{1}'m_{2}'m_{3}'$} &|[block]|{$m_{2}'$} & |[block]|{$m_{2}'\otimes_{p}A_{2}=c_{2}$} &|[block]|{$A_{2}$}  &    |[block]| {$A_{1},A_{2},A_{3}$ }                                         \\
		&  |[block]|{$m_{3}'$} &|[block]|{$m_{3}'\otimes_{p}A_{3}=c_{3}$}  &|[block]| {$A_{3}$ }  &                                        \\                                        
	};
	\draw [line width=0.2mm,>=latex,->] (m-3-1) -- (m-3-2);
	\draw[line width=0.2mm,>=latex,->] (m-3-1) -- (m-2-2);
	\draw [line width=0.2mm,>=latex,->] (m-3-1) -- (m-4-2);
	\path [line width=0.2mm,>=latex,->] (m-3-1) edge (m-3-2);
	\path [line width=0.2mm,>=latex,->] (m-2-2) edge (m-2-3);
	\path [line width=0.2mm,>=latex,->] (m-2-4) edge (m-2-3);
	\path [line width=0.2mm,>=latex,->] (m-3-2) edge (m-3-3);
	\path [line width=0.2mm,>=latex,->] (m-3-4) edge (m-3-3);
	\path [line width=0.2mm,>=latex,->] (m-4-2) edge (m-4-3);
	\path [line width=0.2mm,>=latex,->] (m-4-4) edge (m-4-3);
	\draw [line width=0.2mm,>=latex,->] (m-3-5) -- (m-2-4);
	\draw [line width=0.2mm,>=latex,->] (m-3-5) -- (m-3-4);
	\draw [line width=0.2mm,>=latex,->] (m-3-5) -- (m-4-4);
	
	\end{tikzpicture}
\end{figure} 

\subsection{Decryption Algorithm}
Since the initial vector $I_1$, non-singular linear transformation $T$ and invertible matrix $A_1$ are known at the receiver (say Bob) end,  he  calculates all the required keys and whitening vectors. To get the $i^{\text{th}}$ plaintext block, following operation is needed to perform by the receiver: $$m_i=\big(c_i\otimes_p A_i^{-1}\big)\ominus_pI_i,$$ where $\ominus_p$ represents subtraction modulo $p$.
\subsection{Rationale behind choosing $T$ }
We have chosen the  non-singular linear transformation $T$ over the vector space of dimension $n$ having large order $t$ (by order, we mean the order of $T$ in $ GL(n,\mathbb{F}_p)$ ). If this is not the case, then the key matrices may start repeating themselves before encryption of $ t-1 $ blocks. To avoid this, we have chosen the non-singular transformation of large order.
\section{Security Analysis}
In this section, we discuss the security analysis of the proposed scheme. 
Our encryption scheme is such that the  key matrix used to encrypt any whitened message block is invertible. 
In this scheme,  message block with all  zero entries is not encrypted to the same block  because of whitening. Further, change in one entry of initial vector changes the entire ciphertext and this highlights the importance of the initial vector. Brute force complexity and resistance with various attacks are discussed in subsequent subsections.
\subsection{Brute force attack}
Let $V$ be the same $n$-dimensional vector space over $\mathbb{F}_p$ considered in Section $3$. Then number of basis of $V$ are  $$N=(p^n-1)(p^n-p)\cdots (p^n-p^{n-1}).$$ Due to whitening process and dynamic key for each block, brute force complexity of the proposed scheme has enhanced considerably. In order to mount brute force attack an adversary should have knowledge of prime number $p$ and block size $n$. Assuming availability of these two parameters to an adversary he needs a triplet $(I,B,T)$, where $I \in V$ and $B,T\in GL(n, \mathbb{F}_p).$ Thus, the total number of triplets required for brute force attack are $L=p^n\cdot N^2$. If both $p$ and $n$ are large, then this is very large number and makes the brute force infeasible, e.g. if $n=128$ and $p=29$, then $L>2^{877}$ which is cryptographically secure even with the best available computation facility. This shows that brute force complexity of this cipher is greater than that  of the classical Hill cipher.
\subsection{Security against known-plaintext attack}
In classical Hill cipher same matrix is used for encrypting all the blocks of a message. This causes a consequential weakness, as because of this, a system of $n^2$ linear equation with $n^2-$unknown is formed and can be solved uniquely with the help of Gauss elimination method and Cramer's rule, if coefficient matrix is non-singular. Now, we show that this type of approach is not applicable in our proposed scheme.

Suppose the attacker is available with plaintext-ciphertext pairs $(m_j, c_j), \ j \in J$, where $J$ is some index set. Since different keys are used for encrypting different message blocks, so to find the key matrix used for encrypting $j^{\text{th}}$ message block, an attacker can incorporate only one plaintext-ciphertext pair $(m_j, c_j)$ provided corresponding initial vector is known to the attacker which itself is a difficult task. This gives the linear system of $n$ equations with $n^2$ unknowns and solving this system yields infinitely many solutions. Therefore, our proposed scheme thwart the known-plaintext attack and similarly it is secure against choosen-plaintext attack.
\subsection{Completeness effect}
Each letter of ciphertext block in this scheme depends on all letters of corresponding plain text block. So our proposed scheme has completeness effect.
\section{Computational Costs}
The total time taken by a computer for a specific task, fundamentally depends on the number of bits operations used for its execution and further,  number of bits operations depend on the algorithms. But to analyze the worst case scenario, in the proposed scheme, conventional method bit operation complexity is considered which can, however, be further reduced by employing efficient algorithms \cite{knuth1981art}. Now,  whenever we add, divide and multiply two numbers, then the total numbers of bit operations depend on the bit size of these numbers. In our proposed algorithm, every number is between $0$ and $p-1$, but for worst case situation every numbers in the process will be considered of $ \floor*{\log_{2}(p-1)}+1$ bits.

Let $T_{Enc_{t}}$ and $T_{Dec_{t}}$ signify the time taken by the proposed algorithms for encrypting and decrypting $t^{th}$ block (with $t>1$) of the plaintext and ciphertext, respectively. Assuming $(t-1)^{th}$ block encryption matrix and the corresponding initial vector are given, then on neglecting the followings:
\begin{enumerate}
	\item Time required for "bookkeeping" or logical steps other than the bit operations.
	\item Computational cost of introduced protocol,
\end{enumerate}
we have:
\begin{equation}
T_{Enc_{t}}\approx (n^{3}+n^{2}-n)T^{+_{p}}+(n^{3}+2n^{2})T^{\times_{p}},
\end{equation}
\begin{equation}
T_{Dec_{t}}\approx (3n^{3}-n^{2}-n)T^{+_{p}}+(3n^{3}+2n^{2})T^{\times_{p}}+nT^{-1_{p}},
\end{equation} 
where inverse of dynamic key matrices has been calculated using method of Gauss elimination and the time required for addition, multiplication and inverse in the finite field $\mathbb{F}_p$ are denoted by $T^{+_{p}}$, $T^{\times_{p}}$ and $T^{-1_{p}}$, respectively. Let $\lambda$ be number of bits in $p-1$, i.e. $\lambda=\floor*{log_{2}(p-1)}+1$. Then according to conventional method \cite{rosen1988elementary}, we have 
\begin{equation}
T^{+_{p}}={O}(\lambda)
\end{equation}
\begin{equation}
T^{\times_{p}}={O}(\lambda^{2})
\end{equation}
\begin{equation}
T^{-1_{p}}={O}(\lambda^{3})
\end{equation}
Thus, from equations $5.1$ to $5.5$, it is clear that running time for encryption and decryption explicitly depend on $\lambda$ (bit length of $p-1$) and $n$ (block size).

Let $\wp$ denotes plaintext length. Then overall time taken for encryption of the entire plaintext message is 
\begin{equation}
T_{Total\_Enc}\approx n^2 T^{+_{p}}+n^2	T^{\times_{p}}+( \ceil*{ \frac{\wp}{n}}-1)((n^{3}+n^{2}-n)T^{+_{p}}+(n^{3}+2n^{2})	T^{\times_{p}}).
\end{equation}
Similarly, time taken for decryption of the entire ciphertext is:
\begin{equation}
\begin{split}
T_{Total\_{Dec}}\approx n^2(2n-1)T^{+_{p}}+n^2(2n+1)	T^{\times_{p}}+nT^{-1_{p}}+( \ceil*{ \frac{\wp}{n}}-1)\\ \times((3n^{3}-n^{2}-n)T^{+_{p}}+(3n^{3}+2n^{2})T^{\times_{p}}+nT^{-1_{p}}).
\end{split}
\end{equation}
The computational cost for encrypting/decrypting the whole plaintext/ciphertext  message under the proposed scheme can be calculated by employing equations $5.1$ to $5.5$ in equation $5.6$ and $5.7$, respectively. The comparison of operations required for encryption/decryption of each plaintext/ciphertext block between our scheme and the other schemes based on the Hill cipher is given below in Table $1$. It reflects that the number of operations in our proposed scheme are greater than from others schemes based on the Hill cipher.
\section{Example}
In this section, we discuss a toy example for the feel of our scheme. We restrain ourselves to a key matrices of order $3$ and prime number $p=29$, so that the calculations can be performed manually.
\begin{example}
	Let $V=\{(v_1,v_2,v_3): v_i \in \mathbb{F}_{29}~ for~ 1\leq i\leq3\}$. Clearly $F=\mathbb{F}_{29}$ and length of the message block to be encrypted is $3$. Further, consider the non-singular linear transformation $T:V \to V$ defined by $$T(v_1,v_2,v_3)=(v_1+v_2,3v_2+v_3, v_1-v_2+v_3),$$ initial vector $I_1= (2, 1, 5)$ and the  random basis $$B_1=\{(1, 2, 0), (3, 1, 0), (1, 28, 4)\}.$$ Let $m=m_1m_2m_3m_4m_5m_6$, where $$m_1=(12, 0, 17), \ m_2=( 2, 7, 5), \ m_3=(14, 17, 22),$$ $$ m_4=(0, 17, 3), \ m_5=(0, 19, 5), \ m_6=(8, 21, 4)\hspace{5mm}$$ is the message. Initial whitening of the message yields $m_1'm_2'm_3'm_4'm_5'm_6'$, where $$m_1'=(14, 1, 22), \ m_2'=( 5, 15, 11), \ m_3'=(25, 18, 23),$$ $$  m_4'=(12, 21, 14), \ m_5'=( 16, 13, 24), \ m_6'=(18, 22, 16).$$ Dynamic keys for the encryption  are $$A_1=\begin{bmatrix}
	1&2&0\\3&1&0\\1&-1&4
	\end{bmatrix}, \qquad A_2= \begin{bmatrix}
	3&6&-1\\4&3&2\\0&1&6
	\end{bmatrix},\qquad A_3=\begin{bmatrix}
	9&17&-4\\7&11&3\\1&9&5
	\end{bmatrix}\hspace{4mm}$$ $$A_4=\begin{bmatrix}
	26&18&-12\\18&7&-1\\10&3&-3
	\end{bmatrix}, \qquad A_5= \begin{bmatrix}
	15&13&-4\\25&20&10\\13&6&4
	\end{bmatrix},\quad A_6=\begin{bmatrix}
	-1&6&-2\\16&12&15\\19&22&11
	\end{bmatrix}.$$Corresponding ciphertext is $c_1c_2c_3c_4c_5c_6$ with $$c_1=(10, 7, 1), \ c_2=(17, 28, 4), \ c_3=(26, 26, 11),$$ $$ c_4=(18, 28, 25), \ c_5=(7, 3, 17), \ c_6=(0, 28, 6),$$ where all the operations are performed under modulo $29$. Decryption can be performed easily by computing the inverses of key matrices and knowledge of initial vector. 
\end{example}
\vspace{2mm}
\hspace{10.5mm}\textbf{Table $1$. Bit operation comparison per block among the schemes based }
\\
\vspace{1mm}
\hspace{14mm}\textbf{ on the Hill cipher}

\begin{center}
	\begin{tabular}{ |c|c|c|c|c|c|} 
		\hline
		Distinct Schemes & Operation & {$T_{Mul}$} &$T_{Add}$ &$T_{Inv}$ \\
		\hline
		\multirow{2}{6em}{Original~Hill\\ \hspace{6mm}Cipher} & Encryption & $n^2$ &$n^2-1$ & -\\  \cline{2-5}
		& Decryption & $n^2$ & $n^2-1$&- \\ \hline
		\multirow{2}{6em}{Affine~Hill\\ \hspace{4mm}Cipher}	& Encryption &  $n^2$ & $n^2$ & -\\ \cline{2-5}
		& Decryption &  $n^2$ &$n^2$&- \\ \hline
		\multirow{2}{6em}{Lin et al.'s\\ Scheme \cite{lin2004comments}}	& Encryption &  $n^2+n+3$ &$n^2+4$ &- \\ \cline{2-5}
		& Decryption & $n^2+n+3$ & $n^2+4$&1 \\ \hline
		\multirow{2}{7em}{M.~Toorani~and\\ \hspace{0.1mm}A.~Falahati \cite{toorani2009secure}}	& Encryption & $n^2+2n$ &$n^2+n+1$ &- \\ \cline{2-5}
		& Decryption & $n^2+2n$ & $n^2+n+1$&1 \\ \hline
		\multirow{2}{6em}{Proposed\\ \hspace{1.5mm}Scheme} & Encryption~(first block) & $n^2$ &$n^2$ & -\\  \cline{2-5}
		& Encryption~(block$\neq$1) & $n^{3}+2n^{2}$ &$n^{3}+n^{2}-n$ &- \\ \cline{2-5}
		& Decryption~(first block) & $n^2(2n+1)$ &$ n^2(2n-1)$ &$n$ \\ \cline{2-5}
		& Decryption~(block$\neq$1) & $3n^{3}+2n^2$ &$3n^{3}-n^{2}-n$ &$n$ \\ 
		\hline
	\end{tabular}
\end{center}
\section{Conclusion}
In this paper, we have pointed the traditional Hill cipher algorithm and its drawbacks. To overcome the issues, we have introduced a new variant of Hill cipher. More specifically, we propose a vector space-based approach in which dynamic keys are produced with the help of a  non-singular transformation and basis. This scheme is secure against the known-plaintext attack and  a Brute  Force  attack  requires $((p^n-1)(p^n-p)\cdots (p^n-p^{n-1}))^2$ number  of  key pairs  generations other than initial vector or with assumption that initial vector is known to adversary, where $n$  is  the  order  of  key  matrix. Moreover, proposed framework uses linear operations i.e. matrix additions and multiplications which makes it more efficient. 
\bibliographystyle{plain}
\bibliography{hillcipher}
\end{document}